\documentclass[twocolumn,epjc3]{svjour3}  
\RequirePackage[T1]{fontenc}
\smartqed  
\RequirePackage{graphicx}
\usepackage{enumitem}
\usepackage{amsmath}
\usepackage{graphics} 
\journalname{Eur. Phys. J. C}
\begin{document}
\sloppy

\title{Study of $B$, $B_s$ mesons using heavy quark effective theory
}


\author{Keval Gandhi$^{1,a}$, Ajay Kumar Rai$^{1,b}$ 
}

\thankstext{e1}{e-mail: keval.physics@yahoo.com}
\thankstext{e2}{e-mail: raiajayk@gmail.com}


\institute{Department of Physics, Sardar Vallabhbhai National
Institute of Technology, Surat, Gujarat 395007, India\\
 \label{addr1}
}

\date{Received: date / Accepted: date}

\maketitle

\begin{abstract}
Inspired by the lower statistical information in the bottom {\bf sector, in this paper, we} calculate the masses and the strong decays of excited $B$ and $B_s$ mesons in the framework of heavy quark effective theory (HQET). Using an effective chiral Lagrangian approach based on heavy quark spin-flavor and light quark chiral symmetry, we explore the flavor independent parameters $\Delta_F^{(c)} = \Delta_F^{(b)}$ and $\lambda_F^{(c)} = \lambda_F^{(b)}$ to calculate the masses of experimentally unknown bottom mesons. Our predictions are consistent with the available experimental results and other theoretical studies. Their strong decay to the ground state bottom mesons plus light pseudoscalar mesons is calculated in terms of the square of the couplings $g_H$, $g_S$, $g_T$, $g_X$, $g_Y$, and $g_R$. The weighted average value of the couplings $g_H$, $g_S$ and $g_T$ {\bf is} obtained in the charm sector [Phys. Rev. D \textbf{86}, 054024 (2012)] by fitting the calculated decay widths with experimental measurements, which will be used in the present study to analyze the strong decays of excited open bottom mesons. {\bf Moreover, the ratio of the decay rates is} also predicted, which can be countered with future experimental data.
\end{abstract}

\section{Introduction}
\label{sec1}

In the past decade, evidence of the charmed mesons {\bf has} increased rapidly and {\bf remarkably} compared to the bottom mesons \cite{Zyla2020}. Due to large non-resonant continuum contributions, however, experimentally the broad resonance states are difficult to identify. {\bf To date only the} ground state and the low-lying excited states of $B$ and $B_s$ mesons ($B_1(5721)$, $B_J^*(5732)$, $B_2^*(5747)$, $B_{s1}(5830)$, $B_J(5840)$, $B_{s2}^*(5840)$, and $B_J(5970)$) are reported by various experimental groups such as LHCb \cite{aaij2015precise,aaij2013first}, CDF \cite{aaltonen2014study,aaltonen2009measurement}, and D0 \cite{abazov2007properties}. {\bf In the bottom sector,} we believe that more experimental statistics may be reported in the coming years. For that, the LHCb experiment working at the CERN {\bf will be in a unique position}. 

At the latest, LHCb \cite{aaij2020} measured the masses $(M)$ and decay widths $(\Gamma)$ of two new states of excited $B_s$ meson in the $B^+K^-$ decay mode as

\begin{equation}
\begin{aligned}
\label{eq1} 
(M_1, \Gamma_1) =~& (6063.5 \pm 1.2 \pm 0.8, ~ 26 \pm 4 \pm 4) ~ MeV,\\
(M_2, \Gamma_2) =~& (6114 \pm 3 \pm 5, ~ 66 \pm 18 \pm 21) ~ MeV,\\
\end{aligned}
\end{equation}

\noindent with {\bf the} first statistical and second systematic uncertainties. Based on predictions from various theoretical and phenomenological studies for the masses of excited $B_s$ mesons \cite{godfrey2016b,sun2014higher,ebert2010heavy,lu2016excited,kher2017spectroscopy,devlani2012spectroscopy,shah2016spectroscopy,jia2019regge,chen2018regge,lahde2000spectra}, these newly observed states may result {\bf from the first} orbitally excited states. The $B_s(1^3D_3)$ is predicted {\bf with a relatively} narrow decay width of less than 50 MeV \cite{sun2014higher,Xiao2014,Asghar2018}. 
The states $1^1D_2$ and $1^3D_2$ can be a mix, depending on the mixing angle one of the states from {\bf them is as narrow as} 20 MeV. The spectroscopic notation, $\cal{N}$ $^{2S+1}L_J$, is commonly used to describe these states; where $\cal{N}$ is the radial quantum number, $S$ is the sum of the intrinsic spin of the quarks, $L$ is the orbital angular momentum of the quarks{\bf, and} $J$ gives the total angular momentum. 

The heavy quark effective theory (HQET) is an important tool for calculating properties of heavy-light systems where the heavy quark is assumed to have infinite mass ($m_Q \rightarrow \infty$) \cite{eichten1993properties,mannel1996review,falk1996excited}. In which the total angular momentum of the system is defined as $\vec{J} = \vec{s}_Q + \vec{s}_l$, where $\vec{s}_Q$ is the spin of the heavy quark and $\vec{s}_l =  \vec{s}_{\bar{q}} + \vec{L}$ is the spin of the light degrees of freedom, $s_Q$ and $s_l$ are conserved in the heavy quark mass limit. Here, $\vec{s}_{\bar{q}}$ and $\vec{L}$ are the spin and the orbital angular momentum of the light antiquark, respectively \cite{close1992effective,neubert1992heavy}. For $P$-wave ($L$ = 1), two doublets $\vec{s}_l = {\frac{1}{2}}$ and $\vec{s}_l = {\frac{3}{2}}$, containing four states $J^P_{s_l} = (0^+, 1^+)_{{\frac{1}{2}}}$ and $J^P_{s_l} = (1^+, 2^+)_{{\frac{3}{2}}}$. The first doublet has not been filled experimentally to date, while the second doublet {\bf is filled} with $B_{s1}(5730)^0$ and $B_{s2}^*(5840)^0$, respectively. 

In heavy quark limit, using available experimental information on open charm mesons Ref. \cite{colangelo2012new} {\bf predicts} the masses of some unobserved bottom mesons. The same formalism used in Ref. \cite{gupta2019placing} by Gupta and Upadhyay, and examine the states $B_{s1}(5730)^0$ and $B_{s2}^*(5840)^0$ as the strange partners of $B_J(5721)$ and $B_{s2}^*(5747)$, respectively. In Ref. \cite{gupta2018analysis}, they analyzed experimentally available excited open charmed mesons to {\bf predict similar spectra} in open bottom mesons. At the latest, G.-L. Yu and Z.-G. Wang \cite{yu2019analysis} studied the strong decays of the experimentally observed excited bottom and strange bottom mesons {\bf with the $^3P_0$} decay model. They identify $B_1(5721)$ as $1^3P_1$ and $B_2^*(5747)$ {\bf as $1^3P_2$ and} suggest the quantum states $2^3S_1$ and $1^3D_3$ for $B_J(5840)$ and $B(5970)$ mesons, respectively \cite{yu2019analysis}. These theoretical predictions are not consistent with each other. Therefore, a further test of calculations against the experimental measurements is required to identify the excited $B$ and $B_s$ mesons. 

The main aim of this work is to discuss the possible spin-parity of experimentally observed excited bottom mesons. We shall calculate masses and strong decay behaviors of excited bottom mesons in the framework of leading order approximation of heavy quark effective theory. The flavor symmetry of heavy quark {\bf explores} the flavor independent parameters $\Delta_F^{(c)} = \Delta_F^{(b)}$ and $\lambda_F^{(c)} = \lambda_F^{(b)}$ to calculate the masses of excited excited $B$ and $B_s$ mesons. Their strong two-body decays to ground state bottom mesons plus light pseudoscalar mesons are calculated. The suppression factor arising from the breakdown of isospin symmetry, which occurs when the mass difference between the parent heavy-light meson with strangeness and the daughter nonstrange meson is smaller than the kaon mass, is taken into account here. Our investigation confirmed the spin-parity of experimentally observed open bottom mesons and we believe that this study could provide valuable information in future experimental studies.

This paper is organized as follows: in the framework of heavy quark effective theory section \ref{Theoretical framework} presents the mass parameters of excited heavy-light flavor mesons and {\bf discusses} their strong decays to a member {\bf of the lowest-lying} negative parity states plus light pseudoscalar mesons ($\pi$, $\eta$, and $K$). In section \ref{Mass Spectroscopy: Results and Discussion}, we explore the flavor independent parameters in heavy quark mass limit and calculate the mass spectra of excited open bottom mesons. Their two-body strong decays are analyzed in section \ref{Strong Decay: Numerical Analysis}. Finally, a summary is given in section \ref{Summary}.  

\section{Theoretical framework}
\label{Theoretical framework}

In the leading order approximation, the heavy quark effective theory is developed by expanding the QCD Lagrangian {\bf in the power} of $1/m_Q$ as

{\small{\begin{equation}
\begin{aligned}
\label{eq2}
{\cal{L}}{_{HQET}} = {\cal{L}}{_0} + \frac{1}{m_Q} {\cal{L}}{_1} + \frac{1}{m_Q^2} {\cal{L}}{_2} + ...,
\end{aligned} 
\end{equation}}}

\noindent where finite heavy quark mass corrections are applied and the heavy quark symmetry breaking terms are studied order by order \cite{chen2017review}. Here the leading Lagrangian ${\cal{L}}{_0} = \bar{h}_\nu (i\nu \cdot D)h_\nu$ has an exact spin-flavor symmetry of heavy quark, therefore, only the first term survives in the heavy quark mass limit (a detailed discussion is given in \cite{manohar2000}).

\begin{table*}
\caption{Calculated values of spin-averaged masses $\bar{M}_F$ (in MeV), mass splitting $\Delta_F$ (in MeV), and the hyperfine splitting $\lambda_F$ (in MeV$^2$) of excited $D$ and $D_s$ mesons.}
\begin{tabular*}{\textwidth}{@{\extracolsep{\fill}}ccccccc@{}}
\noalign{\smallskip}\hline\noalign{\smallskip}
&& $D$ meson &&& $D_s$ meson &\\
$\cal{N}$ $J^P_{s_l}$ & $\bar{M}_F$ & $\Delta_F$ & $\lambda_F$ & $\bar{M}_F$ & $\Delta_F$ & $\lambda_F$\\
\noalign{\smallskip}\hline\noalign{\smallskip}
1 $(0^-, 1^-)_{\frac{1}{2}}$ & 1971.3 & $-$ & (262.2)$^2$ &  2076.5 $\pm$ 0.4 & $-$ & (270.2 $\pm$ 0.9)$^2$\\
1 $(0^+, 1^+)_{\frac{1}{2}}$ & 2395.0 $\pm$ 7.3 &  424.0 $\pm$ 7.3 & (240 $\pm$ 24)$^2$ & 2424.3 $\pm$ 0.7  & 347.8 $\pm$ 0.8 & (290.9 $\pm$ 0.8)$^2$\\
1 $(1^+, 2^+)_{\frac{3}{2}}$ & 2448.0 $\pm$ 9.7 & 477.0 $\pm$ 9.7 & (240 $\pm$ 69)$^2$ & 2556.5 $\pm$ 0.2 & 480.0 $\pm$ 0.5 & (255.0 $\pm$ 2.4)$^2$\\
2 $(0^-, 1^-)_{\frac{1}{2}}$ & 2610 $\pm$ 10 & 640 $\pm$ 10 & (200 $\pm$ 28)$^2$ & 2703.0 $\pm$ 2.8 & 627.0 $\pm$ 2.8 & (114.0 $\pm$ 7.9)$^2$ \\
1 $(1^-, 2^-)_{\frac{3}{2}}$ & 2799.2 & 827.9 & (194.3)$^2$ & 2924.2 & 847.7 $\pm$ 0.4 & (198.6)$^2$\\
1 $(2^-, 3^-)_{\frac{5}{2}}$ & 2857.6 & 886.6 & (166.9)$^2$ & 2966.8 & 890.3 $\pm$ 0.4 & (149.2)$^2$\\
2 $(0^+, 1^+)_{\frac{1}{2}}$ & 2928.8 & 957.4 & (97.5)$^2$ & 3063.8 & 987.4 $\pm$ 0.4 & (99.7)$^2$\\
2 $(1^+, 2^+)_{\frac{3}{2}}$ & 2995.0 $\pm$ 4.1 & 1024.0 $\pm$ 4.1 & (300 $\pm$ 25)$^2$ & 3146.5 & 1070.0 $\pm$ 0.4 & (168.32)$^2$\\
3 $(0^-, 1^-)_{\frac{1}{2}}$ & 3087.5 & 1116.2 & (161.8)$^2$ & 3236.2 & 1159.8 $\pm$ 0.4 & (136.3)$^2$\\
1 $(2^+, 3^+)_{\frac{5}{2}}$ & 3112.8 & 1141.4 & (389.3)$^2$ & 3244 &  1167.5 $\pm$ 0.4 & (311.9)$^2$\\
1 $(3^+, 4^+)_{\frac{7}{2}}$ & 3169.5 & 1198.2 & (407.7)$^2$ & 3285.8 & 1209.3 $\pm$ 0.4 & (373.5)$^2$\\
2 $(1^-, 2^-)_{\frac{3}{2}}$ & 3247.4 & 1276 & (274.6)$^2$ & 3395.5 & 1319.0 $\pm$ 0.4 & (225.6)$^2$\\
2 $(2^-, 3^-)_{\frac{5}{2}}$ & 3323.3 & 1352.3 & (264.1)$^2$ & 3463.6 & 1387.0 $\pm$ 0.4 & (183.74)$^2$\\
\noalign{\smallskip}\hline\noalign{\smallskip}
\label{tab1}
\end{tabular*}
\end{table*}

With the help of fields presented in Eq.(10) in Ref. \cite{wang2013analysis}, the kinetic terms of the heavy meson doublets and the field $\Sigma = \xi^2$ of light pseudoscalar mesons are defined in the effective Lagrangian as

{\small{\begin{equation}
\begin{aligned}
\label{eq3} 
\cal{L} = & iTr[\bar{H}_b\nu^{\mu}D_{{\mu}ba}H_a]+\frac{f_{\pi}^{2}}{8}Tr[\partial^{\mu}\Sigma\partial_{\mu}\Sigma^{\dag}] \\
& + Tr[\bar{S}_b(i\nu^{\mu}D_{{\mu}ba}-\delta_{ba}\Delta_S)S_{a}]\\
& + Tr[\bar{T}_b^{\alpha}(i\nu^{\mu}D_{{\mu}ba}-\delta_{ba}\Delta_T)T_{a\alpha}]\\
& + Tr[\bar{X}_b^{\alpha}(i\nu^{\mu}D_{{\mu}ba}-\delta_{ba}\Delta_X)X_{a\alpha}]\\
& + Tr[\bar{Y}_b^{\alpha\beta}(i\nu^{\mu}D_{{\mu}ba}-\delta_{ba}\Delta_Y)Y_{a\alpha\beta}]\\
& + Tr[\bar{Z}_b^{\alpha\beta}(i\nu^{\mu}D_{{\mu}ba}-\delta_{ba}\Delta_Z)Z_{a\alpha\beta}]\\
& + Tr[\bar{R}_b^{\alpha\beta}(i\nu^{\mu}D_{{\mu}ba}-\delta_{ba}\Delta_R)R_{a\alpha\beta}],\\
\end{aligned} 
\end{equation}}}

\noindent where $\Delta_F$ (with $F = S, T, X, Y, Z, R)$ is the mass parameter that gives the mass splittings between the excited and the low-lying negative parity doublets \cite{colangelo2012new,gupta2019placing}. That is written in the form of spin-averaged masses as 
{\small{\begin{equation}
\label{eq4}
\Delta_F = \bar{M}_F - \bar{M}_H,
\end{equation}}}
\noindent with
{\small{\begin{equation}
\begin{aligned}
\label{eq5} 
\bar{M}_H = \frac{3 M_{P^*} + M_P}{4} \hspace{2.25cm}\\
\bar{M}_S = \frac{3 M_{P_{1}^{\prime}} + M_{P_0^*}}{4} \hspace{1cm} \bar{M}_T = \frac{5 M_{P_2^*} + 3 M_{P_1}}{8}\\
\bar{M}_X = \frac{5 M_{P_{2}} + 3 M_{P_1^*}}{8} \hspace{1cm} \bar{M}_Y = \frac{7 M_{P_3^*} + 5 M_{P_2^{\prime}}}{12}\\
\bar{M}_Z = \frac{7 M_{P_{3}} + 5 M_{P_2^{\prime*}}}{12} \hspace{1cm} \bar{M}_R = \frac{7 M_{P_4^*} + 5 M_{P_3^{\prime}}}{12},
\end{aligned} 
\end{equation}}}

\noindent where the ground state doublet $J^P_{s_l} = (0^-, 1^-)_{{\frac{1}{2}}}$ represented by $(P, P^*)$; $P$-wave doublets $J^P_{s_l} = (0^+, 1^+)_{{\frac{1}{2}}}$ and $J^P_{s_l} = (1^+, 2^+)_{{\frac{3}{2}}}$ represented by $(P^*_0, P^{\prime}_1)$ and $(P_1, P_2^*)$, respectively; $D$-wave doublets $J^P_{s_l} = (1^-, 2^-)_{{\frac{3}{2}}}$ and $J^P_{s_l} = (2^-, 3^-)_{{\frac{5}{2}}}$ are represented by $(P^*_1, P_2)$ and $(P^{\prime}_2, P^*_3)$, respectively; and the $F$-wave doublets $J^P_{s_l} = (2^+, 3^+)_{{\frac{5}{2}}}$ and $J^P_{s_l} = (3^+, 4^+)_{{\frac{7}{2}}}$ are represented by $(P^{\prime*}_2, P_3)$ and $(P^{\prime}_3, P^*_4)$, respectively. The mass degeneracy between the members of the meson doublets {\bf is} broken by the Lagrangian as
{\small{\begin{equation}
\begin{aligned}
\label{eq6} 
{\cal{L}}_{{1}/{m_Q}} = & \frac{1}{2m_Q}\{\lambda_HTr[\bar{H}_a\sigma^{\mu\nu}H_a\sigma_{\mu\nu}]-\lambda_STr[\bar{S}_a\sigma^{\mu\nu}S_a\sigma_{\mu\nu}]\\ 
& +\lambda_TTr[\bar{T}_a^{\alpha}\sigma^{\mu\nu}T_{a}^{\alpha}\sigma_{\mu\nu}]-\lambda_XTr[\bar{X}_a^{\alpha}\sigma^{\mu\nu}X_{a}^{\alpha}\sigma_{\mu\nu}]\\
& +\lambda_YTr[\bar{Y}_a^{\alpha\beta}\sigma^{\mu\nu}Y_{a}^{\alpha\beta}\sigma_{\mu\nu}]-\lambda_ZTr[\bar{Z}_a^{\alpha\beta}\sigma^{\mu\nu}Z_{a}^{\alpha\beta}\sigma_{\mu\nu}]\\
& +\lambda_RTr[\bar{R}_a^{\alpha\beta}\sigma^{\mu\nu}R_{a}^{\alpha\beta}\sigma_{\mu\nu}]\}.
\end{aligned} 
\end{equation}}}

The induced symmetry breaking terms are suppressed {\bf by} increasing the powers of inverse heavy quark mass \cite{colangelo2012new,gupta2019placing}. Here the constants $\lambda_H$, $\lambda_S$, $ \lambda_T$, $\lambda_X$, $\lambda_Y$, $\lambda_Z$, and $\lambda_R$ represents the hyperfine mass splitting between the members of the doublets, given by,
{\small{\begin{equation}
\begin{aligned}
\label{eq7} 
\lambda_H = \frac{1}{8}(M^2_{P^*}-M^2_P)  \hspace{2.25cm}\\
\lambda_S = \frac{1}{8}(M^2_{P_1^{\prime}}-M^2_{P_0^*}) \hspace{1cm} \lambda_T = \frac{3}{8}(M^2_{P_2^*}-M^2_{P_1})\\
\lambda_X = \frac{3}{8}(M^2_{P_2}-M^2_{P_1^*}) \hspace{1cm} \lambda_Y = \frac{3}{8}(M^2_{P_3^*}-M^2_{P_2^{\prime}})\\
\lambda_Z = \frac{5}{8}(M^2_{P_3}-M^2_{P_2^{\prime*}}) \hspace{1cm} \lambda_R = \frac{5}{8}(M^2_{P_4^*}-M^2_{P_3^{\prime}}).\\
\end{aligned} 
\end{equation}}}

Flavor symmetry implies that the mass splitting $\Delta_F$ between doublets and the mass splitting $\lambda_F$ between spin-partners in doublets are free from the heavy quark flavor, $i.e.$, 
\begin{equation}
\label{eq8}
\Delta_F^{(c)} = \Delta_F^{(b)} \hspace{1cm} \lambda_F^{(c)} = \lambda_F^{(b)};
\end{equation}
where $F$ stands for $H, S, T, X, Y, Z,$ and $R$.

\begin{table*}
\caption{\label{tab2} Predicted masses of excited $B$ and $B_s$ mesons (in MeV).}
\begin{tabular*}{\textwidth}{@{\extracolsep{\fill}}ccccccccc@{}}
\hline
&& $B$ meson &&&& $B_s$ meson &&\\
$\cal{N}$ $^{2S+1}L_J$ & Present & Exp. \cite{Zyla2020} & Ref. \cite{godfrey2016b} & Ref. \cite{sun2014higher} & Present & Exp. \cite{Zyla2020} & Ref. \cite{godfrey2016b} & Ref. \cite{sun2014higher} \\
\hline
$1^3P_0$ & 5700 $\pm$ 10 & 5698 $\pm$ 8 & 5756 & 5756 & 5709.0 $\pm$ 4.4 & & 5831 & 5830\\
& & $B_J^*(5732)$ & & & & & &\\
$1^1P_1$ & 5740 $\pm$ 11 & 5726.1 $\pm$ 1.3 & 5777 & 5779 & 5768.0 $\pm$ 3.1 & 5828.7 $\pm$ 0.2 & 5857 & 5858\\
& & $B_1(5721)^0$ & & & & $B_{s1}(5730)^0$ & & \\
$1^3P_1$ & 5780 $\pm$ 20 & & 5784 & 5782 & 5875.0 $\pm$ 2.4 & & 5861 & 5859 \\
$1^3P_2$ & 5800 $\pm$ 16 & 5739.5 $\pm$ 0.7 & 5797 & 5796 & 5890.0 $\pm$ 1.7 & 5839.9 $\pm$ 0.1 & 5876 & 5875\\
& & $B_2^*(5747)^0$ & & & & $B_{s2}^*(5840)^0$ & &\\
$2^1S_0$ & 5930 $\pm$ 12 & 5863 $\pm$ 9 & 5904 & 5904 & 6025.0 $\pm$ 3.2 & & 5984 & 5985\\
& & $B_J(5840)^0$ & & & & & &\\
$2^3S_1$ & 5960 $\pm$ 11 & 5971 $\pm$ 5 & 5933 & 5934 & 6033.0 $\pm$ 2.4 & 6063.5 $\pm$ 1.2 & 6012 & 6013\\
& & $B_J(5970)^0$ & & & & $B_s(6063)$ & &\\
$1^3D_1$ & 6136.3 $\pm$ 0.4 & & 6110 & 6110 & 6247.0 $\pm$ 2.4 & & 6182 & 6181\\
$1^1D_2$ & 6144.5 $\pm$ 0.3 & & 6095 & 6108 & 6256.0 $\pm$ 1.7 & & 6169 & 6180\\
$1^3D_2$ & 6196.3 $\pm$ 0.6 & & 6124 & 6113 & 6292.0 $\pm$ 3.5 & & 6196 & 6185\\
$1^3D_3$ & 6202.3 $\pm$ 0.4 & & 6106 & 6105 & 6297.0 $\pm$ 2.5 & & 6179 & 6178\\
$2^3P_0$ & 6266.3 $\pm$ 0.2 & & 6213 & 6214 & 6387.0 $\pm$ 1.3 & & 6279 & 6279\\
$2^1P_1$ & 6272.4 $\pm$ 0.2 & & 6197 & 6206 & 6393.0 $\pm$ 0.9 & & 6279 & 6284\\
$2^3P_1$ & 6325.0 $\pm$ 8.1 & & 6228 & 6219 & 6470.0 $\pm$ 2.4 & & 6296 & 6291\\
$2^3P_2$ & 6344.0 $\pm$ 6.5 & & 6213 & 6213 & 6476.0 $\pm$ 1.7 & & 6295 & 6295\\
$3^1S_0$ & 6417.4 $\pm$ 0.2 & & 6335 & 6334 & 6556.0 $\pm$ 1.3 & & 6410 & 6409\\
$3^3S_1$ & 6433.7 $\pm$ 0.2 & & 6335 & 6355 & 6567.3 $\pm$ 0.9 & & 6429 & 6429\\
$1^3F_2$ & 6436.5 $\pm$ 0.6 & & 6387 & 6387 & 6560.0 $\pm$ 3.5 & & 6454 & 6453\\
$1^1F_3$ & 6467.8 $\pm$ 0.4 & & 6358 & 6375 & 6580.0 $\pm$ 2.4 & & 6425 & 6441\\
$1^3F_3$ & 6499.8 $\pm$ 0.6 & & 6396 & 6380 & 6613.0 $\pm$ 3.5 & & 6462 & 6446\\
$1^3F_4$ & 6520.2 $\pm$ 0.4 & & 6364 & 6364 & 6629.0 $\pm$ 2.5 & & 6432 & 6431\\
$2^3D_1$ & 6580.0 $\pm$ 0.4 & & 6475 & 6475 & 6717.0 $\pm$ 2.3 & & 6542 & 6542\\
$2^1D_2$ & 6595.2 $\pm$ 0.3 & & 6450 & 6464 & 6728.0 $\pm$ 1.6 & & 6526 & 6536\\
$2^3D_2$ & 6657.3 $\pm$ 0.6 & & 6486 & 6472 & 6790.0 $\pm$ 3.5 & & 6553 & 6542\\
$2^3D_3$ & 6671.3 $\pm$ 0.4 & & 6460 & 6459 & 6796.0 $\pm$ 2.5 & & 6535 & 6534\\
\hline
\end{tabular*}
\end{table*}

The effective heavy meson chiral Lagrangians (define in Eq.(11) in Ref. \cite{wang2013analysis}) determine the expressions of strong decays of heavy-light mesons to a member of lowest-lying heavy-light spin doublet plus light pseudoscalar mesons ($\pi$, $\eta$, and $K$),

\begin{equation}
\label{eq9}
\Gamma = \frac{1}{2J+1} \sum \frac{\vec{P}_{\cal{P}}}{8\pi P_a^2} |{\cal{M}}|^2;
\end{equation}

For $B_2^*(5747)$ as $1^3P_2$, the two-body strong decays $B_2^*(5747) \rightarrow B^*\pi, B\pi$; having the $\pi$ meson three momenta ${\vec{P}_{\cal{\pi}}} =$ 374 and 418 MeV, respectively. The decay widths    

\begin{equation}
\label{eq10}
\Gamma(B_2^*(5747) \rightarrow B^*\pi, B\pi) \propto {\vec{P}_{\cal{\pi}}^5},
\end{equation}

\noindent where ${\vec{P}_{\cal{\pi}}^5} = $ 7.3 $\times$ 10$^{12}$  and 1.3 $\times$ 10$^{13}$ MeV$^5$ in the {\bf decay} to the final states $B^*\pi$ and $B\pi$, respectively. A small difference in ${\vec{P}_{\cal{\pi}}}$ can lead to a {\bf significant} difference in ${\vec{P}_{\cal{\pi}}^5}$. {\bf So,} we have to take into account the heavy quark symmetry breaking corrections and chiral symmetry breaking corrections to make robust predictions. The higher-order corrections for spin and flavor violation of an order ${\cal{O}}(\frac{1}{m_Q})$ are not taking into consideration to avoid introducing new unknown coupling constants \cite{wang2015}. We expect that the corrections would not be larger than (or as large as) the leading order contributions. At the hadronic level, the $\frac{1}{m_Q}$ corrections can be crudely estimated to be of the order $\frac{{\vec{P}_{\cal{\pi}}}}{M_B} \left(\frac{{\vec{P}_{K}}}{M_{B_s}}\right) \approx$ 0.1-0.2. 

\section{Mass Spectroscopy: Results and Discussion}
\label{Mass Spectroscopy: Results and Discussion}

In our previous study, we analyzed the strong decay of experimentally observed excited $D$ and $D_s$ mesons \cite{gandhi2019strong,gandhi2020identifying}. The ratio of the strong decay rates {\bf has} identified the doublets: $(D_1(2420), D_2^*(2460))$ as $(1^3P_1, 1^3P_2)$, $(D_{s0}^*(2317), D_{s1}(2460))$ as $(1^3P_0, 1^1P_1)$, $(D_{s1}(2536),\\
D_{s2}^*(2573))$ as $(1^3P_1, 1^3P_2)$ $(D(2550), D_J^*(2600))$ as\\
$(2^1S_0, 2^3S_1)$ and $(D(2740), D_J^*(2750))$ as $(1^3D_2, 1^3D_3)$. The resonances $D{_{J}^*}(3000)$ and $D_2^*(3000)$ are identified with $2^3P_2$ and $1^3F_2$ quantum states, respectively. The $D_J(3000)$ and $D_{sJ}(3040)$ {\bf are} interpreted as the mixing of $2^1P_1 - 2^3P_1$ states. 

The calculated values of spin-averaged masses $\bar{M}_F$, the mass splitting between the doublets $\Delta_F$, and the mass splitting between spin-partner in a doublet $\lambda_F$ are listed in Table (\ref{tab1}). {\bf Further, we compute their propagation in the determination of uncertainty using the generic method given in the Appendix. Uncertainty arises in presented results because of the uncertainties in the various experimental inputs.} Here, the masses of $D^0(D_s^{\pm})$ as $(1^1S_0)$, $D^*(2007)^0 (D_s^{*\pm})$ as $(1^3S_0)$, $D_0^*(2400)^0 (D_{s0}^*(2317)^{\pm})$ as $(1^3P_0)$, $D_1(2430)^0 (D_{s1}(2460)^{\pm})$ as $(1^1P_1)$, $D_1(2420)^0\\ (D_{s1}(2536))$ as $(1^3P_1)$, $D_2^*(2460)^0 (D_{s2}^*(2573))$ as $(1^3P_2)$, $D(2550)^0$ as $(2^1S_0)$, $D_J^*(2600) (D_{s1}^*(2700))$ as $(2^3S_1)$, $D_J(3000)$ as $(2^3P_1)$, and $D_J^*(3000)$ as $(2^3P_2)$ are taken from PDG \cite{Zyla2020}. Experimentally the $1D$, $2P$, $1F$, {\bf and} $2D$ states of excited $D$ and $D_s$ mesons are missing (or PDG \cite{Zyla2020} need more confirmation), $i.e.$ the predictions of $D(2740)$ as $(1^3D_2)$, $D_J^*(2750)$ as $(1^3D_3)$, $D_{s1}^*(2860)$ as $(1^3D_1)$, $D_{s3}^*(2860)$ as $(1^3D_3)$, $D_2^*(3000)$ as $(1^3F_2)$, and $D_{sJ}(3040)^{\pm}$ as a mixture of $2P(J=1)$, are still uncertain. {\bf Also,} $D_s(2^1S_0)$, the spin-partner of $D_s(2^3S_1)$ is still missing experimentally. So their masses are taken from the predictions of Ebert, Faustov, and Galkin (EFG) \cite{ebert2010heavy}.

The charm data can be exploited to make predictions of excited bottom mesons. Our calculated masses of the excited $B$ and $B_s$ mesons are listed in Table (\ref{tab2}) with available experimental observations and theoretical predictions. Note that the ground state masses of the $B$ and $B_s$ mesons are fixed from PDG \cite{Zyla2020}. {\bf For $1P$-wave, our calculated mass 5700 $\pm$ 10 MeV of $B(1^3P_0)$ is in good agreement with experimental observed mass 5698 $\pm$ 8 MeV \cite{Zyla2020} of $B_J^*(5732)$. Experimentally $B_1(5721)^0$ is measured with spin-parity $J^P = 1^+$. That is consistent with our prediction of $B(1^1P_1)$ by a mass difference of $\approx$ 14 MeV. The mass 5768.0 $\pm$ 3.1 MeV of $B_s(1^1P_1)$ and 5875.0 $\pm$ 2.4 MeV of $B_s(1^3P_1)$ are underestimated to the experimental measurement 5828.7 $\pm$ 0.2 MeV of $B_{s1}(5730)^0$ \cite{Zyla2020}. And, the mass of $B_s(1^3P_1)$ is consistent with the results of \cite{godfrey2016b,sun2014higher} with a mass difference of 14-15 MeV. Our result 5800 $\pm$ 16 MeV of $B(1^3P_2)$ is overestimated to the PDG fit value 5739.5 $\pm$ 0.7 MeV \cite{Zyla2020} and is in good agreement with the predictions of \cite{godfrey2016b,sun2014higher}. The mass 5930 $\pm$ 12 MeV of $B(2^1S_0)$ is overestimated to the theoretical and experimental results \cite{Zyla2020,godfrey2016b,sun2014higher} with a mass difference of 26-67 MeV. Our predicted mass 5960 $\pm$ 11 MeV of its spin-partner $B(2^3S_1)$ is within the errorbar of PDG fit value 5971 $\pm$ 5 MeV \cite{Zyla2020}. The calculated masses of $B_s(2S)$ states are underestimated to the LHCb measurement 6063.5 $\pm$ MeV \cite{aaij2020} with a mass difference of 30.5-38.5 MeV and overestimated by the theoretical predictions \cite{godfrey2016b,sun2014higher} with a mass difference of 20-49 MeV.} Here, our predicted masses of $2S$ and $1D$ states are inconsistent with the observed mass 6114 $\pm$ 33 MeV \cite{aaij2020} of excited $B_s$ meson. Our results of 1$D$, 2$P$, $3S$, 1$F$, and 2$D$ states of bottom mesons are overestimated to other theoretical outcomes \cite{godfrey2016b,sun2014higher} (shown in Table (\ref{tab2})). The masses of $1F$ states are close to $3S$ multiplets and overlap. Experimentally $3S$ states are more likely to be observed first rather than $1F$ states, although, it is challenging. 


\section{Strong Decay: Numerical Analysis}
\label{Strong Decay: Numerical Analysis}
\subsection{$B$ meson}
\label{$B$ meson}
\subsubsection{$1P$ and $2P$ states}
\label{$1P$ and $2P$ states(B)}

\begin{table*}
\caption{Strong decay widths of excited $B$ and $B_s$ mesons (in MeV).}
\begin{tabular*}{\textwidth}{@{\extracolsep{\fill}}ccccc@{}}
\noalign{\smallskip}\hline\noalign{\smallskip}
$\cal{N}$ $^{2S+1}L_J$ & Decay & Width & Decay & Width\\
& mode & & mode \\
\noalign{\smallskip}\hline\noalign{\smallskip}
$1^3P_0$ & $B^+\pi^-$ & (589.6 $\pm$ 15.6)$g_S^2$ & $B^+K^-$ & $-$\\  
& $B^0\pi^0$ & (295.0 $\pm$ 7.8)$g_S^2$ & $B^0K^0$ & $-$ \\ 
& $B_sK$ & $-$ & $B_s\pi^0$ & (149.6 $\pm$ 3.9)$g_S^2$  $\times$ $10^{-4}$ \\
& $B^0\eta$ & $-$ & $B_s\eta$ & $-$ \\
\noalign{\smallskip}\hline\noalign{\smallskip}
$1^1P_1$ & $B^*\pi^+$ & (568.9 $\pm$ 15.2)$g_S^2$ & $B^+K^-$ & $-$ \\  
& $B^*\pi^0$ & (285.4 $\pm$ 7.6)$g_S^2$ & $B^0K^0$ & $-$  \\ 
& $B{_s^*}K$ & $-$ & $B_s\pi^0$ & (163.6 $\pm$ 4.3)$g_S^2$ $\times$ $10^{-4}$ \\
& $B^*\eta$ & $-$ & $B_s\eta$ & $-$ \\
\noalign{\smallskip}\hline\noalign{\smallskip} 
$1^3P_1$ & $B^*\pi^+$ & (144.2 $\pm$ 34.0)$g_T^2$ & $B^+K^-$ & (7.5 $\pm$ 0.9)$g_T^2$ \\  
& $B^*\pi^0$ & (73.3 $\pm$ 17.2)$g_T^2$ & $B^0K^0$ & (6.4 $\pm$ 0.8)$g_T^2$ \\ 
& $B{_s^*}K$ & $-$ & $B_s\pi^0$ & (77.0 $\pm$ 3.2)$g_T^2$ $\times$ $10^{-4}$\\
& $B^*\eta$ & $-$ & $B_s\eta$ & $-$ \\
\noalign{\smallskip}\hline\noalign{\smallskip} 
$1^3P_2$ & $B^*\pi^+$ & (108.3 $\pm$ 9.4)$g_T^2$ & $B^{*+}K^-$ & (4.5 $\pm$ 0.4)$g_T^2$ \\  
& $B^*\pi^0$ & (55 $\pm$ 10)$g_T^2$ & $B^{*0}K^0$ & (3.8 $\pm$ 0.4)$g_T^2$ \\ 
& $B{_s^*}K$ & $-$ & $B{_s^*}\pi^0$ & (46.2 $\pm$ 1.9)$g_T^2$ $\times$ $10^{-4}$ \\
& $B^*\eta$ & $-$ & $B_s^*\eta$ & $-$ \\
& $B^+\pi^-$ & (115.1 $\pm$ 18.5)$g_T^2$ & $B^+K^-$ & (14.2 $\pm$ 0.9)$g_T^2$ \\  
& $B^0\pi^0$ & (57.3 $\pm$ 9.2)$g_T^2$ & $B^0K^0$ & (12.9 $\pm$ 0.8)$g_T^2$ \\ 
& $B_sK$ & $-$ & $B_s\pi^0$ & (51.7 $\pm$ 1.8)$g_T^2$ $\times$ $10^{-4}$ \\
& $B^0\eta$ & $-$ & $B_s\eta$ & $-$ \\
\noalign{\smallskip}\hline\noalign{\smallskip}
$2^1S_0$ & $B^*\pi^+$ & (1471.7 $\pm$ 96.3)$g_H^2$ & $B^{*+}K^-$ & (847.1 $\pm$ 31.2)$g_H^2$  \\  
& $B^*\pi^0$ & (739.9 $\pm$ 48.2)$g_H^2$ & $B^{*0}K^0$ & (827.1 $\pm$ 31.1)$g_H^2$ \\ 
& $B{_s^*}K$ & (20.4 $\pm$ 6.7)$g_H^2$ & $B_s^*\pi^0$ & (757.0 $\pm$ 23.8)$g_H^2$ $\times$ $10^{-4}$\\
& $B^*\eta$ & (16.3 $\pm$ 18.0)$g_H^2$ & $B_s^*\eta$ & (90.7 $\pm$ 8.2)$g_H^2$ \\
\noalign{\smallskip}\hline\noalign{\smallskip} 
$2^3S_1$ & $B^*\pi^+$ & (1129.3 $\pm$ 65.7)$g_H^2$ & $B^{*+}K^-$ & (602.0 $\pm$ 11.6)$g_H^2$\\  
& $B^*\pi^0$ & (567.4 $\pm$ 32.9)$g_H^2$ & $B^{*0}K^0$ & (588.3 $\pm$ 11.5)$g_H^2$ \\ 
& $B{_s^*}K$ & (52.7 $\pm$ 19.6)$g_H^2$ & $B_s^*\pi^0$ & (524.2 $\pm$ 7.2) $g_H^2$ $\times$ $10^{-4}$\\
& $B^*\eta$ & (26.2 $\pm$ 5.3)$g_H^2$ & $B_s^*\eta$ & (73.2 $\pm$ 4.7)$g_H^2$\\
& $B^+\pi^-$ & (687.2 $\pm$ 37.6)$g_H^2$ & $B^+K^-$ & (415.8 $\pm$ 6.8)$g_H^2$\\  
& $B^0\pi^0$ & (344.5 $\pm$ 18.8)$g_H^2$ & $B^0K^0$ & (407.3 $\pm$ 6.7)$g_H^2$ \\ 
& $B_sK$ & (81.1 $\pm$ 15.2)$g_H^2$ & $B_s\pi^0$ & (325.8 $\pm$ 3.5)$g_H^2$ $\times$ $10^{-4}$\\
& $B^0\eta$ & (25.3 $\pm$ 3.5)$g_H^2$ & $B_s\eta$ & (84.9 $\pm$ 2.8)$g_H^2$ \\
\noalign{\smallskip}\hline\noalign{\smallskip}
$1^3D_1$ & $B^*\pi^+$ & (869.2 $\pm$ 22.7)$g_X^2$ & $B^{*+}K^-$ & (1017.2 $\pm$ 26.6)$g_X^2$ \\  
& $B^*\pi^0$ & (435.8 $\pm$ 11.4)$g_X^2$ & $B^{*0}K^0$ & (1008.04 $\pm$ 26.3)$g_X^2$ \\ 
& $B{_s^*}K$ & (211.3 $\pm$ 5.5)$g_X^2$ & $B_s^*\pi^0$ & (490.4 $\pm$ 12.8)$g_X^2$  $\times$ $10^{-4}$\\
& $B^*\eta$ & (64.4 $\pm$ 1.7)$g_X^2$ & $B_s^*\eta$ & (305.6 $\pm$ 8.0)$g_X^2$ \\
& $B^+\pi^-$ & (2228.8 $\pm$ 58.2)$g_X^2$ & $B^+K^-$ & (2650.4 $\pm$ 69.2)$g_X^2$  \\  
& $B^0\pi^0$ & (1114.8 $\pm$ 29.1)$g_X^2$ & $B^0K^0$ & (2624.2 $\pm$ 68.5)$g_X^2$ \\
& $B_sK$ & (665.8 $\pm$ 17.4)$g_X^2$ & $B_s\pi^0$ & (1272.1 $\pm$ 33.2)$g_X^2$ $\times$ $10^{-4}$\\ 
& $B^0\eta$ & (185.0 $\pm$ 4.8)$g_X^2$ & $B_s\eta$ & (945.9 $\pm$ 24.7)$g_X^2$ \\
\noalign{\smallskip}\hline\noalign{\smallskip}
$1^1D_2$ & $B^*\pi^+$ & (2732.2 $\pm$ 71.4)$g_X^2$ & $B^{*+}K^-$ & (3223.6 $\pm$ 84.2)$g_X^2$ \\  
& $B^*\pi^0$ & (1369.7 $\pm$ 35.8)$g_X^2$ & $B^{*0}K^0$ & (3195.5 $\pm$ 83.4)$g_X^2$ \\ 
& $B{_s^*}K$ & (688 $\pm$ 18)$g_X^2$ & $B_s^*\pi^0$ & (1546.4 $\pm$ 40.4)$g_X^2$ $\times$ $10^{-4}$\\
& $B^*\eta$ & (207.1 $\pm$ 5.4)$g_X^2$ & $B_s^*\eta$ & (985.9 $\pm$ 25.8)$g_X^2$ \\
\noalign{\smallskip}\hline\noalign{\smallskip}
$1^3D_2$ & $B^*\pi^+$ & (900.7 $\pm$ 23.9)$g_Y^2$ & $B^{*+}K^-$ & (331.5 $\pm$ 13.8)$g_Y^2$ \\  
& $B^*\pi^0$ & (453 $\pm$ 12)$g_Y^2$ & $B^{*0}K^0$ & (324.9 $\pm$ 13.6)$g_Y^2$ \\ 
& $B{_s^*}K$ & (78.4 $\pm$ 2.8)$g_Y^2$ & $B_s^*\pi^0$ & (236.9 $\pm$ 9.3)$g_Y^2$ $\times$ $10^{-4}$ \\
& $B^*\eta$ & (28.1 $\pm$ 0.8)$g_Y^2$ & $B_s^*\eta$ & (60.1 $\pm$ 3.3)$g_Y^2$ \\
\noalign{\smallskip}\hline\noalign{\smallskip}   
$1^3D_3$ & $B^*\pi^+$ & (538.4 $\pm$ 14.2)$g_Y^2$ & $B^{*+}K^-$ & (396.4 $\pm$ 13.8)$g_Y^2$ \\  
& $B^*\pi^0$ & (270.8 $\pm$ 7.1)$g_Y^2$ & $B^{*0}K^0$ & (388.7 $\pm$ 13.6)$g_Y^2$ \\ 
& $B{_s^*}K$ & (48.8 $\pm$ 1.7)$g_Y^2$ & $B_s^*\pi^0$ & (281.0 $\pm$ 9.6)$g_Y^2$ $\times$ $10^{-4}$  \\
& $B^*\eta$ & (17.3 $\pm$ 0.5)$g_Y^2$ & $B_s^*\eta$ & (73.1 $\pm$ 3.3)$g_Y^2$ \\  
\noalign{\smallskip}\hline\noalign{\smallskip}   
\end{tabular*}\\
{continued...}
\label{tab3}
\end{table*}

\begin{table*}
\addtocounter{table}{-1}
\caption{Strong decay widths of excited $B$ and $B_s$ mesons (in MeV).}
\begin{tabular*}{\textwidth}{@{\extracolsep{\fill}}ccccc@{}}
\noalign{\smallskip}\hline\noalign{\smallskip}
$\cal{N}$ $^{2S+1}L_J$ & Decay & Width & Decay & Width\\
& mode & & mode \\
\noalign{\smallskip}\hline\noalign{\smallskip}
& $B^+\pi^-$ & (559.4 $\pm$ 14.7)$g_Y^2$ & $B^+K^-$ & (438.4 $\pm$ 14.8)$g_Y^2$ \\  
& $B^0\pi^0$ & (280.5 $\pm$ 7.4)$g_Y^2$ & $B^0K^0$ & (429.5 $\pm$ 14.5)$g_Y^2$ \\
& $B_sK$ & (69.4 $\pm$ 1.9)$g_Y^2$ & $B_s\pi^0$ & (298.6 $\pm$ 9.5)$g_Y^2$ $\times$ $10^{-4}$  \\
& $B^0\eta$ & (21.9 $\pm$ 0.6)$g_Y^2$ & $B_s\eta$ & (95.9 $\pm$ 3.6)$g_Y^2$\\
\noalign{\smallskip}\hline\noalign{\smallskip}
$2^3P_0$ & $B^+\pi^-$ & (5907.0 $\pm$ 154.2)$g_S^2$ & $B^+K^-$ & (7460.6 $\pm$ 194.8)$g_S^2$ \\  
& $B^0\pi^0$ & (2951.86 $\pm$ 77.1)$g_S^2$ & $B^0K^0$ & (7443.11 $\pm$ 194.4)$g_S^2$ \\ 
& $B_sK$ & (4080.7 $\pm$ 106.6)$g_S^2$ & $B_s\pi^0$ & (3243.1 $\pm$ 84.6)$g_S^2$ $\times$ $10^{-4}$  \\
& $B^0\eta$ & (866.5 $\pm$ 22.6)$g_S^2$ & $B_s\eta$ & (3850.9 $\pm$ 100.6)$g_S^2$ \\
\noalign{\smallskip}\hline\noalign{\smallskip}
$2^1P_1$ & $B^*\pi^+$ & (5319.5 $\pm$ 138.9)$g_S^2$ & $B^{*+}K^-$ & (6757.6 $\pm$ 176.5)$g_S^2$ \\  
& $B^*\pi^0$ & (2661.1 $\pm$ 69.5)$g_S^2$ & $B^{*0}K^0$ & (6746.9 $\pm$ 176.2)$g_S^2$ \\ 
& $B{_s^*}K$ & (3519.2 $\pm$ 91.9)$g_S^2$ & $B_s^*\pi^0$ & (2904.4 $\pm$ 75.9)$g_S^2$ $\times$ $10^{-4}$ \\
& $B^*\eta$ & (768.4 $\pm$ 20.1)$g_S^2$ & $B_s^*\eta$ & (3394.1 $\pm$ 88.6)$g_S^2$ \\
\noalign{\smallskip}\hline\noalign{\smallskip}
$2^3P_1$ & $B^*\pi^+$ & (6653.4 $\pm$ 309.0)$g_T^2$ & $B^{*+}K^-$ & (7606.9 $\pm$ 218.9)$g_T^2$\\  
& $B^*\pi^0$ & (3337.5 $\pm$ 154.8)$g_T^2$ & $B^{*0}K^0$ & (7537.2 $\pm$ 217.0) $g_T^2$ \\ 
& $B{_s^*}K$ & (1874.9 $\pm$ 125.6)$g_T^2$ & $B_s^*\pi^0$ & (4278.2 $\pm$ 124.0)$g_T^2$ $\times$ $10^{-4}$  \\
& $B^*\eta$ & (475.1 $\pm$ 28.9)$g_T^2$ & $B_s^*\eta$ & (2692.6 $\pm$ 84.4)$g_T^2$ \\
\noalign{\smallskip}\hline\noalign{\smallskip}
$2^3P_2$ & $B^*\pi^+$ & (4350.8 $\pm$ 173.7)$g_T^2$ & $B^{*+}K^-$ & (4698.6 $\pm$ 129.1)$g_T^2$ \\  
& $B^*\pi^0$ & (2182.2 $\pm$ 87.0)$g_T^2$ & $B^{*0}K^0$ & (4656.1 $\pm$ 127.9)$g_T^2$ \\ 
& $B{_s^*}K$ & (1289.4 $\pm$ 70.4)$g_T^2$ & $B_s^*\pi^0$ & (2634.0 $\pm$ 73.7)$g_T^2$ $\times$ $10^{-4}$ \\
& $B^*\eta$ & (322.5 $\pm$ 16.1)$g_T^2$ & $B_s^*\eta$ & (1674.4 $\pm$ 49.4)$g_T^2$\\
& $B^+\pi^-$ & (3521.6 $\pm$ 136.8)$g_T^2$ & $B^+K^-$ & (3854.0 $\pm$ 105.3)$g_T^2$ \\  
& $B^0\pi^0$ & (1763.0 $\pm$ 68.5)$g_T^2$ & $B^0K^0$ & (3815.8 $\pm$ 104.3)$g_T^2$ \\ 
& $B_sK$ & (1187.8 $\pm$ 59.3)$g_T^2$& $B_s\pi^0$ & (2143.6 $\pm$ 58.1)$g_T^2$ $\times$ $10^{-4}$  \\
& $B^0\eta$ & (282.0 $\pm$ 13.3)$g_T^2$ & $B_s\eta$ & (1466.9 $\pm$ 40.8)$g_T^2$ \\
\noalign{\smallskip}\hline\noalign{\smallskip}
$3^1S_0$ & $B^*\pi^+$ & (7591.0 $\pm$ 198.3)$g_H^2$ & $B^{*+}K^-$ & (8120.4 $\pm$ 214.0) $g_H^2$ \\  
& $B^*\pi^0$ & (3801.7 $\pm$ 99.3)$g_H^2$ & $B^{*0}K^0$ & (8082.7 $\pm$ 213.0)$g_H^2$\\ 
& $B{_s^*}K$ & (4067.4 $\pm$ 108.7)$g_H^2$ & $B_s^*\pi^0$ & (4282.8 $\pm$ 113.8)$g_H^2$ $\times$ $10^{-4}$  \\
& $B^*\eta$ & (837.4 $\pm$ 21.9)$g_H^2$ & $B_s^*\eta$ & (3922.6 $\pm$ 24.9)$g_H^2$\\
\noalign{\smallskip}\hline\noalign{\smallskip}
$3^3S_1$ & $B^*\pi^+$ & (5262.79 $\pm$ 137.5)$g_H^2$ & $B^{*+}K^-$ & (5570.4 $\pm$ 13.9)$g_H^2$ \\  
& $B^*\pi^0$ & (2635.5 $\pm$ 68.8)$g_H^2$ & $B^{*0}K^0$ & (5545.2 $\pm$ 13.9)$g_H^2$ \\ 
& $B{_s^*}K$ & (2873.4 $\pm$ 76.7)$g_H^2$ & $B_s^*\pi^0$ & (2930.3 $\pm$ 12.4)$g_H^2$ $\times$ $10^{-4}$  \\
& $B^*\eta$ & (588.7 $\pm$ 15.4)$g_H^2$ & $B_s^*\eta$ & (2706.3 $\pm$ 14.9)$g_H^2$ \\
& $B^+\pi^-$ & (2914.44 $\pm$ 76.1)$g_H^2$ & $B^+K^-$ & (3099.9 $\pm$ 7.5)$g_H^2$  \\  
& $B^0\pi^0$ & (1458.1 $\pm$ 38.1)$g_H^2$ & $B^0K^0$ & (3084.3 $\pm$ 7.3)$g_H^2$ \\ 
& $B_sK$ & (1687.6 $\pm$ 44.1)$g_H^2$ & $B_s\pi^0$ & (1627.6 $\pm$ 3.5)$g_H^2$ $\times$ $10^{-4}$ \\
& $B^0\eta$ & (337.3 $\pm$ 8.8)$g_H^2$ & $B_s\eta$ & (1553.5 $\pm$ 4.2)$g_H^2$\\
\noalign{\smallskip}\hline\noalign{\smallskip}
$1^3F_2$ & $B^*\pi^+$ & (1781.5 $\pm$ 46.5)$g_Z^2$ & $B^{*+}K^-$ & (2378.7 $\pm$ 62.1)$g_Z^2$ \\  
& $B^*\pi^0$ & (892.9 $\pm$ 23.3)$g_Z^2$ & $B^{*0}K^0$ & (2361.7 $\pm$ 61.7)$g_Z^2$  \\ 
& $B{_s^*}K$ & (573.9 $\pm$ 15.0)$g_Z^2$ & $B_s^*\pi^0$ & (1084.7 $\pm$ 28.3)$g_Z^2$ $\times$ $10^{-4}$ \\
& $B^*\eta$ & (160.0 $\pm$ 4.1)$g_Z^2$ & $B_s^*\eta$ & (809.5 $\pm$ 21.1)$g_Z^2$ \\
& $B^+\pi^-$ & (3423.6 $\pm$  89.4)$g_Z^2$ & $B^+K^-$ & (4578.5 $\pm$ 118.6)$g_Z^2$ \\  
& $B^0\pi^0$ & (1712.1 $\pm$ 44.7)$g_Z^2$ & $B^0K^0$ & (4539.3 $\pm$ 119.6)$g_Z^2$   \\ 
& $B_sK$ & (1229.8 $\pm$ 32.1)$g_Z^2$ & $B_s\pi^0$ & (2106.0 $\pm$ 55.0)$g_Z^2$ $\times$ $10^{-4}$ \\
& $B^0\eta$ & (324.5 $\pm$ 8.5)$g_Z^2$ & $B_s\eta$ & (1660.2 $\pm$ 43.4)$g_Z^2$ \\
\noalign{\smallskip}\hline\noalign{\smallskip}
$1^1F_3$ & $B^*\pi^+$ & (5312.0 $\pm$ 138.7)$g_Z^2$ & $B^{*+}K^-$ & (6666.4 $\pm$ 174.1)$g_Z^2$ \\  
& $B^*\pi^0$ & (2662 $\pm$ 70)$g_Z^2$ & $B^{*0}K^0$ & (6620.7 $\pm$ 172.9)$g_Z^2$ \\ 
& $B{_s^*}K$ & (1815.0 $\pm$ 47.4)$g_Z^2$ & $B_s^*\pi^0$ & (3026.3 $\pm$ 79.0)$g_Z^2$ $\times$ $10^{-4}$  \\
& $B^*\eta$ & (496.3 $\pm$ 13.0)$g_Z^2$ & $B_s^*\eta$ & (2311.9 $\pm$ 60.4)$g_Z^2$ \\
\noalign{\smallskip}\hline\noalign{\smallskip}
$1^3F_3$ & $B^*\pi^+$ & (27012.7 $\pm$ 716.0)$g_R^2$ & $B^{*+}K^-$ & (29351.3 $\pm$ 1093.9)$g_R^2$ \\  
& $B^*\pi^0$ & (13563.3 $\pm$ 359.5)$g_R^2$ & $B^{*0}K^0$ & (28985.8 $\pm$ 1081.9)$g_R^2$ \\ 
& $B{_s^*}K$ & (5196.0 $\pm$ 158.6)$g_R^2$ & $B_s^*\pi^0$ & (16050.7 $\pm$ 598.2)$g_R^2$ $\times$ $10^{-4}$ \\
& $B^*\eta$ & (1578.6 $\pm$ 42.2)$g_R^2$ & $B_s^*\eta$ & (7805.3 $\pm$ 333.4)$g_R^2$ \\
\noalign{\smallskip}\hline\noalign{\smallskip}
\end{tabular*}\\
{continued...}
\label{tab3}
\end{table*}

\begin{table*}
\addtocounter{table}{-1}
\caption{Strong decay widths of excited $B$ and $B_s$ mesons (in MeV).}
\begin{tabular*}{\textwidth}{@{\extracolsep{\fill}}ccccc@{}}
\noalign{\smallskip}\hline\noalign{\smallskip}
$\cal{N}$ $^{2S+1}L_J$ & Decay & Width & Decay & Width\\
& mode & & mode \\
\noalign{\smallskip}\hline\noalign{\smallskip}
$1^3F_4$ & $B^*\pi^+$ & (17284.0 $\pm$ 454.9)$g_R^2$ & $B^{*+}K^-$ & (18345.7 $\pm$ 589.2)$g_R^2$ \\  
& $B^*\pi^0$ & (8677.2 $\pm$ 228.4)$g_R^2$ & $B^{*0}K^0$ & (18123.6 $\pm$ 582.6)$g_R^2$ \\ 
& $B{_s^*}K$ & (3510.5 $\pm$ 105.3)$g_R^2$ & $B_s^*\pi^0$ & (9942.0 $\pm$ 327.1)$g_R^2$ $\times$ $10^{-4}$  \\
& $B^*\eta$ & (1051.2 $\pm$ 27.8)$g_R^2$ & $B_s^*\eta$ & (4980.1 $\pm$ 181.0)$g_R^2$ \\
& $B^+\pi^-$ & (18610.8 $\pm$ 491.5)$g_R^2$ & $B^+K^-$ & (20108.9 $\pm$ 635.9)$g_R^2$ \\  
& $B^0\pi^0$ & (9282.8 $\pm$ 244.1)$g_R^2$ & $B^0K^0$ & (19832.6 $\pm$ 627.4)$g_R^2$ \\ 
& $B_sK$ & (4352.6 $\pm$ 115.0)$g_R^2$ & $B_s\pi^0$ & (10891.5 $\pm$ 336.7)$g_R^2$ $\times$ $10^{-4}$  \\
& $B^0\eta$ & (1223.3 $\pm$ 32.3)$g_R^2$ & $B_s\eta$ & (5924.5 $\pm$ 196.6)$g_R^2$ \\
\noalign{\smallskip}\hline\noalign{\smallskip}
$2^3D_1$ & $B^*\pi^+$ & (6272.98 $\pm$ 163.8)$g_X^2$ & $B^{*+}K^-$ & (8404.7 $\pm$ 219.5)$g_X^2$ \\  
& $B^*\pi^0$ & (3139.82 $\pm$ 82.0)$g_X^2$ & $B^{*0}K^0$ & (8379.2 $\pm$ 218.8)$g_X^2$\\ 
& $B{_s^*}K$ & (3528.4 $\pm$ 92.1)$g_X^2$ & $B_s^*\pi^0$ & (3717.4 $\pm$ 97.0)$g_X^2$ $\times$ $10^{-4}$ \\
& $B^*\eta$ & (802.8 $\pm$ 21.0)$g_X^2$ & $B_s^*\eta$ & (3883.5 $\pm$ 101.4)$g_X^2$ \\
& $B^+\pi^-$ & (14595.9 $\pm$ 381.1)$g_X^2$ & $B^+K^-$ & (19445.5 $\pm$ 507.8)$g_X^2$ \\  
& $B^0\pi^0$ & (7296.1 $\pm$ 190.5)$g_X^2$ & $B^0K^0$ & (19367.9 $\pm$ 505.8)$g_X^2$ \\ 
& $B_sK$ & (8630.2 $\pm$ 225.4)$g_X^2$ & $B_s\pi^0$ & (8693.1 $\pm$ 227.0)$g_X^2$ $\times$ $10^{-4}$ \\
& $B^0\eta$ & (1906.0 $\pm$ 49.8)$g_X^2$ & $B_s\eta$ & (9269.46 $\pm$ 242.1)$g_X^2$ \\
\noalign{\smallskip}\hline\noalign{\smallskip}
$2^1D_2$ & $B^*\pi^+$ & (19845.8 $\pm$ 518.2)$g_X^2$ & $B^{*+}K^-$ & (26170.2 $\pm$ 683.4)$g_X^2$ \\  
& $B^*\pi^0$ & (9933.2 $\pm$ 259.4)$g_X^2$ & $B^{*0}K^0$ & (26092.3 $\pm$ 681.4)$g_X^2$ \\ 
& $B{_s^*}K$ & (11300.9 $\pm$ 295.1)$g_X^2$& $B_s^*\pi^0$ & (11573.4 $\pm$ 302.2)$g_X^2$ $\times$ $10^{-4}$ \\
& $B^*\eta$ & (2558.9 $\pm$ 66.8)$g_X^2$ & $B_s^*\eta$ & (12148.3 $\pm$ 317.3)$g_X^2$ \\
\noalign{\smallskip}\hline\noalign{\smallskip}
$2^3D_2$ & $B^*\pi^+$ & (13738.2 $\pm$ 361.2)$g_Y^2$ & $B^{*+}K^-$ & (8350.3 $\pm$ 261.6)$g_Y^2$  \\  
& $B^*\pi^0$ & (6886.6 $\pm$ 181.1)$g_Y^2$ & $B^{*0}K^0$ & (8289.8 $\pm$ 259.9)$g_Y^2$ \\ 
& $B{_s^*}K$ & (5026.2 $\pm$ 140.4)$g_Y^2$ & $B_s^*\pi^0$ & (4230.8 $\pm$ 133.7)$g_Y^2$ $\times$ $10^{-4}$ \\
& $B^*\eta$ & (1236.3 $\pm$ 32.6)$g_Y^2$ & $B_s^*\eta$ & (3163.0 $\pm$ 106.5)$g_Y^2$ \\
\noalign{\smallskip}\hline\noalign{\smallskip}
$2^3D_3$ & $B^*\pi^+$ & (8380.4 $\pm$ 219.6)$g_Y^2$ & $B^{*+}K^-$ & (9820.6 $\pm$ 283.6)$g_Y^2$\\  
& $B^*\pi^0$ & (4200.6 $\pm$ 110.1)$g_Y^2$ & $B^{*0}K^0$ & (9750.1 $\pm$ 281.6)$g_Y^2$ \\ 
& $B{_s^*}K$ & (3125.8 $\pm$ 86.8)$g_Y^2$ & $B_s^*\pi^0$ & (4968.4 $\pm$ 146.1)$g_Y^2$ $\times$ $10^{-4}$  \\
& $B^*\eta$ & (764.7 $\pm$ 20.1)$g_Y^2$ & $B_s^*\eta$ & (3734.8 $\pm$ 114.7)$g_Y^2$ \\
& $B^+\pi^-$ & (7674.2 $\pm$ 201.2)$g_Y^2$& $B^+K^-$ & (9033.7 $\pm$ 259.0)$g_Y^2$ \\  
& $B^0\pi^0$ & (3839.8 $\pm$ 100.6)$g_Y^2$ & $B^0K^0$ & (8958.1 $\pm$ 256.9)$g_Y^2$  \\ 
& $B_sK$ & (3094.9 $\pm$ 81.2)$g_Y^2$ & $B_s\pi^0$ & (4592.0 $\pm$ 130.6)$g_Y^2$ $\times$ $10^{-4}$ \\
& $B^0\eta$ & (728.8 $\pm$ 19.1)$g_Y^2$ & $B_s\eta$ & (3597.9 $\pm$ 105.4)$g_Y^2$\\
\noalign{\smallskip}\hline\noalign{\smallskip}
\end{tabular*}
\label{tab3}
\end{table*}

Our predicted mass of $B(1^3P_0)$ is above {\bf the} $B\pi$ and $B^*\pi$ threshold, but due to the parity conservation{\bf, only decay} in $B\pi$ mode is possible. Using the weighted average value of coupling $g_S$ = 0.56 (obtained in the strong decay analysis of the $1P$-wave doublet $(D_0^*(2400),\\ D_1(2420)) = (1^3P_0, 1^1P_1)$) from \cite{colangelo2012new}, {\bf we get the partial decay width of $B(1^3P_0) \rightarrow B\pi$ = 278 $\pm$ 3 MeV. That is slightly overestimated to the QCD sum rule prediction in HQET 250 MeV \cite{zhu1999effect}, and consistent with the chiral quark model value 272 MeV \cite{zhong2008strong}, and larger than the results of quark pair creation (QPC) model 230 MeV \cite{lu2016excited} and 225 MeV \cite{sun2014higher}}. Such a broad decay width is the fundamental reason for the experimental unavailability of this state. For its spin-partner $B(1^1P_1)$, taking a sum of its partial decay width listed in Table (\ref{tab3}), {\bf which reach up to 267.0 $\pm$ 5.3 MeV. This state is very broad compared to the $B_1(5721)$ measurements 30.1 $\pm$ 1.5 MeV of LHCb \cite{aaij2015precise} and 23 $\pm$ 3 MeV of CDF \cite{aaltonen2014study}}. From the strong decay analysis of $B(1^1P_1)$, we conclude that $B_1(5721)$ is not a member of doublet $(0^+, 1^+)_{{\frac{1}{2}}^+}$.

According to HQET, there are two $1^+$ states in $P$-wave doublets $(0^+, 1^+)_{{\frac{1}{2}}^+}$ and $(1^+, 2^+)_{{\frac{3}{2}}^+}$ with $\vec{s}_l^P = {\frac{1}{2}}^+$ and $\vec{s}_l^P = {\frac{3}{2}}^+$, respectively. {\bf For the second doublet $(1^+, 2^+)_{{\frac{3}{2}}^+}$, the decay widths 41 $\pm$ 7 MeV $B(1^3P_1)$ and 63 $\pm$ 4.6 MeV $B(1^3P_2)$ are calculated by taking a sum of its partial decay widths listed in Table (\ref{tab3}) and using the weighted average value of coupling $g_T$ = 0.43 (obtained in the strong decay analysis of the $1P$-wave doublet $(D_1(2430), D_2^*(2460)) = (1^3P_1, 1^3P_2)$ and $(D_{s2}^*(2573)) = (1^3P_2)$) from \cite{colangelo2012new}. For $B(1^3P_1)$, the decay width 41 $\pm$ 7 MeV is slightly overestimated to LHCb \cite{aaij2015precise} and CDF \cite{aaltonen2014study} measurements 30.1 $\pm$ 1.5 MeV and 23 $\pm$ 3 MeV, respectively. For natural parity state $B(1^3P_2)$, the decay width 63 $\pm$ 4.6 MeV is overestimated to the experimentally observed narrow decay width of $B_{2}^*(5747)$: 24.5 $\pm$ 1.0 MeV of LHCb \cite{aaij2015precise} and 22$^{+3}_{-2}$ MeV of CDF \cite{aaltonen2014study}, and the theoretical result 47 MeV of \cite{zhong2008strong}.}

The ratio of the decay rates is independent of the unknown hadronic couplings and compare with the experimental results where available. The ratio,
\begin{equation} 
\label{eq11} 
\frac{\Gamma(B(1^3P_2) \rightarrow B^*\pi)}{\Gamma(B(1^3P_2) \rightarrow B\pi)} = 1.0 \pm 1.1, 
\end{equation}
\noindent {\bf is within the errorbar of LHCb \cite{aaij2015precise} measurements 0.71 $\pm$ 0.14 $\pm$ 0.30 and 0.91 $\pm$ 0.13 $\pm$ 0.12 for neutral and charged states respectively, and the result of $D0$ experiment 1.10 $\pm$ 0.42 $\pm$ 0.31 \cite{abazov2007properties}. The best estimate value 1.0 (of Eq. (\ref{eq11})) is close to the other theoretical result 0.9 of \cite{sun2014higher,lu2016excited,gupta2019placing,zhong2008strong}.} Also, we find out the ratio    
\begin{align} 
\label{eq12} 
R_1 &= \frac{\Gamma(B(1^3P_1)\rightarrow B^*\pi))}{\Gamma(B(1^3P_1)\rightarrow B^*\pi)) + \Gamma(B(1^3P_2)\rightarrow B^*\pi))} \\
&= 0.6 \pm 0.9,
\end{align}
\noindent and
\begin{align} 
\label{eq13} 
R_2 &= \frac{\Gamma(B(1^3P_2)\rightarrow B^*\pi))}{\Gamma(B(1^3P_2)\rightarrow B^*\pi)) + \Gamma(B(1^3P_2)\rightarrow B\pi))} \\
& = 0.6 \pm 0.7,
\end{align} 
\noindent {\bf which are consistent with the $D0$ \cite{abazov2007properties} measurements $R_1$ = 0.47 $\pm$ 0.06 and $R_2$ = 0.47 $\pm$ 0.09, and underestimated to the theoretical prediction $R_1$ = 0.3 and consistent with $R_2$ = 0.5 of \cite{zhong2008strong}, and close to $R_1$ = 0.6 and $R_2$ = 0.5 of \cite{gupta2019placing}.} So, we write 
\begin{equation}
\label{eq14} 
B_{J}^*(5732) = (0^+)_{{\frac{1}{2}}^+} = \big(1^3P_0\big).
\end{equation}
Its spin-partner $B(1^1P_1)$ is still not found experimentally. Its identification is an important task in future experimental studies. The doublet,
\begin{equation}
\label{eq15} 
\big(B_1(5721), B_{2}^*(5747)\big) = (1^+, 2^+)_{{\frac{3}{2}}^+} = \big(1^3P_1, 1^3P_2\big),
\end{equation}
\noindent is identified very well in the present study. Since the experimental information of the $2P$ states is missing. For the $2P$ states, more decay channels are open. Suppose the strong decays of excited $B$ meson (listed in Table (\ref{tab3})) are contributing dominantly to the total decay width,  then we can give an information of fundamental decay modes of the particular decay state. The $B(2^3P_0)$ dominantly {\bf decays} in $B^+\pi^-$ and $B_sK$ modes, and $B^*\pi^+$ and $B^+\pi^-$ are the main decay channels of $B(2^3P_2)$. The $B(2^1P_1)$ and $B(2^3P_1)$ states can be found in $B^*\pi^+$ decay mode.

\subsubsection{$2S$ and $3S$ states}
\label{$2S$ and $3S$ states(B)}

The experimentally observed mass of $B_{J}(5840)$ is close to the $B(2^1S_0)$ and $B(2^3S_1)$ assignments. The $B_{J}(5840)$ was observed in $B\pi$ decay mode, i.e. the member of natural parity states. Therefore, we can exclude the $B(2^1S_0)$ assignment of $B_{J}(5840)$. Using the weighted average value of coupling $g_H$ = 0.28 (obtained in the strong decay analysis of the $1P$-wave doublet $(D_1(2550),\\ D_J^*(2600)) = (2^1S_0, 2^3S_1)$ and $(D_{s1}^*(2700)) = (2^3S_1)$) from \cite{colangelo2012new} and taking the sum of the partial decay widths of $B(2^3S_1)$ presented in Table (\ref{tab3}), {\bf which reach up to 229.0 $\pm$ 6.9 MeV. This state is very broad and in good agreement with the LHCb \cite{aaij2015precise} measurement 224.4 $\pm$ 23.9 MeV of $B_{J}(5840)$ and overestimated to the theoretical results 107.8 MeV of \cite{godfrey2016b}, 106.1 MeV of \cite{lu2016excited}, and 121.9 MeV of \cite{yu2019analysis}.} Therefore, the strong decay analysis suggests $B(2^3S_1)$ for $B_{J}(5840)$, but still uncertain. It can also be a candidate of $B(1^3D_1)$ and $B(1^3D_3)$ states. The calculated mass of the $B(2^3S_1)$ state is close to $B(1^3D_1)$ and $B(1^3D_3)$ states. {\bf However,} experimentally the $B(2^3S_1)$ states might be more likely to be observed than the $1D$ states. So, we write 
\begin{equation}
\label{eq16} 
\big(B_{J}(5840)\big) = (1^-)_{{\frac{1}{2}}^-}  = \big(2^3S_1\big).
\end{equation}
Our predicted decay width 229.0 $\pm$ 6.9 MeV of $B(2^3S_1)$ state with coupling $g_H$ = 0.28 is much larger compare to LHCb \cite{aaij2015precise} measurement 82 $\pm$ 8 MeV of $B_{J}(5970)$. Also, the mass difference $ M_{B_{J}(5970)} - M_{B_{J}(5840)}$ $\approx$ 100 MeV. So that they may not be members of the same wave family. The $B^*\pi$ is the main decay channel of $B(2^3S_1)$, its branching ratio relative to $B\pi$ decay mode is 
\begin{equation}
\label{eq17} 
\frac{\Gamma(B(2^3S_1) \rightarrow B^*\pi)}{\Gamma(B(2^3S_1) \rightarrow B\pi)} = 1.6 \pm 0.1,
\end{equation}
\noindent calculated from Table (\ref{tab3}), 
{\bf which is underestimated to the predictions 1.9 of \cite{godfrey2016b}, 2.5 of \cite{sun2014higher}, and 2.3 of \cite{lu2016excited}.} From Table (\ref{tab3}), the $3S$ states are dominantly decaying in $B^*\pi^+$ mode {\bf compared} to $B^+\pi^-$. For $B(3^3S_1)$, the branching ratio 
\begin{equation}
\label{eq18} 
\frac{\Gamma(B(3^3S_1) \rightarrow B^*\pi)}{\Gamma(B(3^3S_1) \rightarrow B\pi)} = 1.8 \pm 0.1,
\end{equation}
\noindent {\bf is predicted in the range of other theoretical estimations 1.4 of \cite{godfrey2016b} and 2.4 of \cite{sun2014higher}.} 

\subsubsection{$1D$, $2D$, and $1F$ states}
\label{$1D$, $2D$, and $1F$ states(B)}  

There are four $1D$ states, from which an unnatural parity states $B(1^1D_2)$ and $B(1^3D_2)$ are mainly to decay in $B^*\pi$ mode (see Table (\ref{tab3})), and in accordance with the results of Refs. \cite{godfrey2016b,sun2014higher,lu2016excited}. For the natural parity states $B(1^3D_1)$ and $B(1^3D_3)$, the branching ratio
\begin{equation}
\label{eq19} 
R_3 = \frac{\Gamma(B(1^3D_1) \rightarrow B\pi)}{\Gamma(B(1^3D_1) \rightarrow B^*\pi)} = 2.6 \pm 0.1,
\end{equation}
\noindent and
\begin{equation}
\label{eq20} 
R_4 = \frac{\Gamma(B(1^3D_3) \rightarrow B\pi)}{\Gamma(B(1^3D_3) \rightarrow B^*\pi)} \approx 1.0,
\end{equation}
\noindent are calculated from Table (\ref{tab3}). {\bf The value of  $R_3$ is overestimated to 2.0 of \cite{godfrey2016b,sun2014higher}, 1.7 of \cite{lu2016excited}; while $R_4$ is consistent with 1.0 of \cite{godfrey2016b,sun2014higher,lu2016excited}.} Therefore, the $B\pi$ mode is more dominant over $B^*\pi$ for $B(1^3D_1)$ state than $B(1^3D_3)$. $B_{J}(5970)$ is observed in the decay of two pseudoscalar mesons. Thus, we suggest
\begin{equation}
\label{eq21} 
\big(B_{J}(5970)\big) = (1^-)_{{\frac{3}{2}}^-}  = \big(1^3D_1\big).
\end{equation}
{\bf We obtain the coupling $g_X$ $\sim$ 0.1} by equating the calculated decay width of $B(1^3D_1)$ from Table (\ref{tab3}) with LHCb \cite{aaij2015precise} and CDF \cite{aaltonen2014study} measurements for $B_{J}(5970)$. No experimental information is available for $2D$ and $1F$ states. From the results listed in Table (\ref{tab3}), the $B(2^3D_1)$ is mainly to decay in $B\pi$, and the $2^3D_3$ state is equally found in $B\pi$ and $B^*\pi$ modes. The branching ratio,
\begin{equation}
\label{eq22} 
R_5 = \frac{\Gamma(B(2^3D_1) \rightarrow B\pi)}{\Gamma(B(2^3D_1) \rightarrow B^*\pi)} = 2.3 \pm 0.9,
\end{equation}
\noindent and
\begin{equation}
\label{eq23} 
R_6 = \frac{\Gamma(B(2^3D_3) \rightarrow B\pi)}{\Gamma(B(2^3D_3) \rightarrow B^*\pi)} \approx 1.0.
\end{equation}
\noindent {\bf The $R_5$ is consistent with 2.5 of \cite{godfrey2016b} and 2.2 of \cite{sun2014higher}, and for $R_6$ our prediction is larger than 0.5 of \cite{godfrey2016b}.} The decay behavior of the $1F$ states is shown in Table (\ref{tab3}). The $BK$ mode is dominant over $B^*K$, for the $B(1^3F_2)$ and $B(1^3F_4)$ states. The branching ratio,
\begin{equation}
\label{eq24} 
R_7 = \frac{\Gamma(B(1^3F_2) \rightarrow B\pi)}{\Gamma(B(1^3F_2) \rightarrow B^*\pi)} = 1.9 \pm 0.1,
\end{equation}
\noindent and
\begin{equation}
\label{eq25} 
R_8 = \frac{\Gamma(B(1^3F_4) \rightarrow B\pi)}{\Gamma(B(1^3F_4) \rightarrow B^*\pi)} \approx 1.1.
\end{equation}
\noindent {\bf Here, the $R_7$ is quite larger than 1.4 of \cite{godfrey2016b,sun2014higher}, and $R_8$ is consistent with 1.0 of \cite{godfrey2016b}.} $B^*\pi$ is the main decay channel for $B(2^1D_2)$, $B(2^3D_2)$, $B(1^1F_3)$, $B(1^3F_3)$ states and in accordance with the results of Ref. \cite{godfrey2016b} and \cite{sun2014higher}. The ratios of the strong decay rates will be helpful {\bf in searching} these states experimentally.

\subsection{$B_s$ meson}
\label{$B_s$ meson}
\subsubsection{$1P$ and $2P$ states}
\label{$1P$ and $2P$ states(Bs)}

The predicted mass of $B_s(1^3P_0)$ is below $BK$ threshold \cite{lu2016excited,zeng1995heavy,cheng2014near,di2001excited}. A {\bf similar situation arises,} like in the strange charmed partner $D_{s0}^*(2317)$. Hence, only the strong decay to $B_s\pi$ mode (called isospin-breaking mode \cite{gross1979}) is allowed kinematically. To account for isospin violation, the square of the suppression factor, given by,  
\begin{equation}
\label{eq26} 
\epsilon^2 = \frac{3}{16} \left(\frac{m_d-m_u}{m_s-\frac{m_u+m_d}{2}}\right)^2 \simeq 10^{-4},
\end{equation}

\noindent is multiplied with the decay width formula \cite{colangelo2003understanding,matsuki2012}. Here $m_u$, $m_d$, and $m_s$ are the current quark masses. Using the weighted average value of coupling $g_S$ = 0.56 (obtained in the strong decay analysis of the $1P$-wave doublet $(D_0^*(2400), D_1(2420)) = (1^3P_0, 1^1P_1)$) from \cite{colangelo2012new}, {\bf we get the partial decay width of $B_s(1^3P_0) \rightarrow B_s\pi^0$ = (47.0 $\pm$ 1.2) $\times$ $10^{-4}$ MeV. And, its doublet partner $B_s(1^1P_1) \rightarrow B_s\pi^0$ = (51.0 $\pm$ 1.3) $\times$ $10^{-4}$ MeV.} Here the decay $B_s(1^3P_0)$ $\rightarrow$ $BK$ is kinematically forbidden. S. Godfrey et al. \cite{godfrey2016b} used the QPC model and determined the partial decay width 138 MeV of $B_s(1^3P_0)$ decay to $BK$ mode. Such a broad decay width of $B_s(1^3P_0)$ following some other QPC model predictions 225 MeV of \cite{sun2014higher} and 217 MeV of \cite{yu2019analysis}, and the result from the chiral quark model 227 MeV \cite{zhong2008strong}. Such a broad resonance state is difficult to identify in experimental studies.

The ratio among the decay rates can be used to confirm or reject the quantum number assignments of the observed states. The ratio,
\begin{equation} 
\label{eq27} 
\frac{\Gamma(B_s(1^3P_2) \rightarrow B^{*+}K^-)}{\Gamma(B_s(1^3P_2) \rightarrow B^{+}K^-)} \approx 0.3, 
\end{equation}
\noindent is calculated using the results of Table (\ref{tab3}). {\bf It is larger than the experimental measurements 0.081 $\pm$ 0.021 $\pm$ 0.015 of CMS \cite{collaboration2018studies}, 0.093 $\pm$ 0.013 $\pm$ 0.012 of LHCb \cite{aaij2013first} and 0.10 $\pm$ 0.03 $\pm$ 0.02 of CDF \cite{aaltonen2008observation}, and the theoretical prediction 0.1 of \cite{sun2014higher,lu2016excited,yu2019analysis,zhong2008strong}. The $B_{s2}^*(5840)$ as $B_s(1^3P_2)$, with PDG \cite{Zyla2020} world average mass 5839.92 $\pm$ 0.14 MeV, our calculated partial decay rates are (0.38 $\pm$ 0.02)$g_T^2$ MeV and (4.63 $\pm$ 0.13)$g_T^2$ MeV for the decay states $B^{*+}K^-$ and $B^{+}K^-$, respectively.} That gives a ratio $\frac{\Gamma({B^{*+}K^-})}{\Gamma({B^{+}K^-})}$ $\approx$ 0.082 $\pm$ 0.005, which is within the errorbar of CMS \cite{collaboration2018studies} measurement, and in agreement with the results of other experimental and theoretical studies \cite{aaij2013first,sun2014higher,yu2019analysis,zhong2008strong,aaltonen2008observation}.

{\bf The decay width 6.7 $\pm$ 0.3 MeV of $B_s(1^3P_2)$ is calculated} by taking a sum of its partial decay width listed in Table (\ref{tab3}) and using the weighted average value of coupling $g_T$ = 0.43 (obtained in the strong decay analysis of the $1P$-wave doublet $(D_1(2430), D_2^*(2460)) = (1^3P_1, 1^3P_2)$ and $(D_{s2}^*(2573)) = (1^3P_2)$) from \cite{colangelo2012new}. {\bf Such a narrow decay width of $B_s(1^3P_2)$ is quite larger than the experimental observed narrow resonance state $B_{s2}^*(5840)$: 1.52 $\pm$ 0.34 $\pm$ 30 MeV of CMS \cite{collaboration2018studies}, 1.4 $\pm$ 0.4 $\pm$ 0.2 MeV of LHCb \cite{aaij2013first}, and 1.56 $\pm$ 0.13 $\pm$ 0.47 MeV of CDF \cite{aaltonen2008observation}, and the theoretical prediction 2 MeV of \cite{lu2016excited,zhong2008strong}. Therefore, $B_{s2}^*(5840)$ can be a strong candidate {\bf for} $B_s(1^3P_2)$. For its doublet partner $B_s(1^3P_1)$, the decay width 2.6 $\pm$ 0.2 MeV is obtained by summing up the partial decay width from Table (\ref{tab3}) with $g_T$ = 0.43. Such a state is narrow and is underestimated to the prediction 21.4 MeV of \cite{lu2016excited} and overestimated to the CDF measurement 0.5 $\pm$ 0.3 $\pm$ 0.3 MeV \cite{aaltonen2014study}.} In $P$-wave, the states $1^1P_1$ and $1^3P_1$ can be a mix, and as we said in the {\bf introduction,} one of the states may narrow. Experimentally $B_{s1}(5830)$ is observed as a narrow state \cite{aaltonen2014study}. So it is more reliable to predict $B_s(1^3P_1)$. Therefore, we write
\begin{equation}
\label{eq28} 
\big(B_{s1}(5830), B_{s2}^*(5840)\big) = (1^+, 2^+)_{{\frac{3}{2}}^+} = \big(1^3P_1, 1^3P_2\big).
\end{equation}
For the $2P$ states, more decay channels are open. But there has been no experimental observation to date. Suppose the strong decays of excited strange-bottom mesons (listed in Table (\ref{tab3})) are contributing dominantly to the total decay {\bf width, we} can give information of {\bf the} fundamental decay modes of the particular decay state. The $BK$ is the main decay mode of $B_s(2^3P_0)$, and the states $B_s(2^1P_1)$ and $B_s(2^3P_1)$ are dominantly decaying in $B^*K$ mode; and $B^*K$ and $BK$ {\bf are} the main decay channels of $B_s(2^3P_2)$. Our results are consistent with the predictions of \cite{godfrey2016b,sun2014higher}.

\subsubsection{$2S$ and $3S$ states}
\label{$2S$ and $3S$ states(Bs)}

The decay behavior of the $B_s(2^1S_0)$ and $B_s(2^3S_1)$ states are shown in Table (\ref{tab3}). It shows that the $B^*K$ is the main decay mode of $B_s(2^1S_0)$, and for $B_s(2^3S_1)$ state $B^*K$ is dominant over $BK$. The branching ratio,
\begin{equation}
\label{eq29} 
\frac{\Gamma(B_s(2^3S_1) \rightarrow B^*K)}{\Gamma(B_s(2^3S_1) \rightarrow BK)} \approx 1.4,
\end{equation}
\noindent {\bf which is in good agreement with 1.5 of \cite{godfrey2016b}, and overestimated to 2.0 of \cite{sun2014higher,lu2016excited}.} {\bf Like} $2S$ states, the $B^*K$ is the main decay mode of $B_s(3^1S_0)$ state. {\bf Moreover}, for $B_s(3^3S_1)$, the $B^*K$ mode is dominant over $BK$. The branching ratio, 
\begin{equation}
\label{eq30} 
\frac{\Gamma(B_s(3^3S_1) \rightarrow B^*K)}{\Gamma(B_s(3^3S_1) \rightarrow BK)} \approx 1.8,
\end{equation}
\noindent {\bf is predicted in the range of the theoretical results 1.2 of \cite{godfrey2016b} and 2.0 of \cite{sun2014higher}.} It is interesting to study the strong decays of {\bf a} strange-bottom meson with the help of couplings {\bf obtained in the charm sector}. Using the weighted average value of coupling $g_H$ = 0.28 (obtained in the strong decay analysis of the $1P$-wave doublet $(D_1(2550), D_J^*(2600)) = (2^1S_0, 2^3S_1)$ and $(D_{s1}^*(2700)) = (2^3S_1)$) from \cite{colangelo2012new} and taking the sum of the partial decay widths of $B_s(2^1S_0)$ and $B_s(2^3S_1)$ presented in Table (\ref{tab3}), which {\bf reach} up to {\bf 138.0 $\pm$ 3.5 MeV and 170.0 $\pm$ 1.5 MeV, respectively. Such states are predicted as a broad resonance state, which is underestimated to the predictions 213.4 MeV $B_s(2^1S_0)$ and 221.9 MeV $B_s(2^3S_1)$ of Ref. \cite{lu2016excited}, and overestimated to 75.8 MeV $B_s(2^1S_0)$ and 114 MeV $B_s(2^3S_1)$ of Ref. \cite{godfrey2016b}, and 44 MeV $B_s(2^1S_0)$ and 51 MeV $B_s(2^3S_1)$ of Ref. \cite{sun2014higher}.} 

\subsubsection{$1D$, $2D$, and $1F$ states}
\label{$1D$, $2D$, and $1F$ states(Bs)}

The natural parity $1D$ states $1^3D_1$ and $1^3D_3$ are dominant to decay in $BK$ (see Table (\ref {tab3})). Their calculated branching ratio of the decay rates relative to $B^*K$ decay mode are,
\begin{equation}
\label{eq31} 
R_1 = \frac{\Gamma(B_s(1^3D_1) \rightarrow BK)}{\Gamma(B_s(1^3D_1) \rightarrow B^*K)} = 2.6 \pm 0.1,
\end{equation}
\noindent and
\begin{equation}
\label{eq32} 
R_2 = \frac{\Gamma(B_s(1^3D_3) \rightarrow BK)}{\Gamma(B_s(1^3D_3) \rightarrow B^*K)} \approx 1.1.
\end{equation}
\noindent {\bf Our calculated value of $R_1$ is overestimated by the theoretical results 2.0 of \cite{godfrey2016b}, 2.2 of \cite{sun2014higher}, and 1.8 of \cite{lu2016excited}; and $R_2$ is consistent with 1.2 of \cite{godfrey2016b}, 0.9 of \cite{sun2014higher}, and 1.1 of \cite{lu2016excited}.} The $B^*K$ is the main channel of $B_s(1^1D_2)$ and $B_s(1^3D_2)$, which are in accordance with the results of Refs. \cite{godfrey2016b,sun2014higher,lu2016excited}. For $2^3D_1$ states, $BK$ is the fundamental decay mode; and for $2^3D_3$, $B^*K$ is dominant over $BK$ (see Table (\ref{tab3})). The branching ratio,  

\begin{equation}
\label{eq33} 
R_3 = \frac{\Gamma(B_s(2^3D_1) \rightarrow BK)}{\Gamma(B_s(2^3D_1) \rightarrow B^*K)} = 2.3 \pm 0.1,
\end{equation}
\noindent and
\begin{equation}
\label{eq34} 
R_4 = \frac{\Gamma(B_s(2^3D_3) \rightarrow BK)}{\Gamma(B_s(2^3D_3) \rightarrow B^*K)} \approx 0.9.
\end{equation}
\noindent Here, $R_3$ is consistent with 2.5 of \cite{godfrey2016b} and 2.3 of \cite{sun2014higher}, while for $R_4$ our prediction is quite larger than 0.5 of \cite{godfrey2016b} and 0.4 of \cite{sun2014higher}. For $B_s(2^1D_2)$ and $B_s(2^3D_2)$ states, $B^*K$ is the main decay channel. The decay behavior of the $1F$ states is shown in Table (\ref{tab3}). The $BK$ mode is dominant over $B^*K$, for the states $B_s(1^3F_2)$ and $B_s(1^3F_4)$. The branching ratio,  
\begin{equation}
\label{eq35} 
R_5 = \frac{\Gamma(B_s(1^3F_2) \rightarrow BK)}{\Gamma(B_s(1^3F_2) \rightarrow B^*K)} = 1.9 \pm 0.1,
\end{equation}
\noindent and
\begin{equation}
\label{eq36} 
R_6 = \frac{\Gamma(B_s(1^3F_4) \rightarrow BK)}{\Gamma(B_s(1^3F_4) \rightarrow B^*K)} \approx 1.1.
\end{equation}
\noindent {\bf Here, the $R_5$ is overestimated to 1.4 of \cite{godfrey2016b,sun2014higher}, and $R_6$ is consistent with 1.0 of \cite{godfrey2016b}.} The branching ratio distinguishes the fundamental decay mode, which is valuable to further experimental search. {\bf Using $g_X$ $\sim$ 0.1 (as we extract in sec \ref{$1D$, $2D$, and $1F$ states(B)} for $B_{J}(5970)$ as $B(1^3D_1)$) and taking summation of partial decay rates of $B_s(1^3D_1)$ and $B_s(1^1D_2)$ listed in Table (\ref{tab3}), we obtained 82.0 $\pm$ 1.0 MeV and 74.0 $\pm$ 1.2 MeV, respectively. These are underestimated to the results 213.4 MeV $B_s(1^3D_1)$ and 198.6 MeV $B_s(1^1D_2)$ of \cite{lu2016excited}, 183 MeV $B_s(1^3D_1)$ of \cite{godfrey2016b}, and 191 MeV $B_s(1^3D_1)$ of \cite{sun2014higher}. Our calculated decay width 82.0 $\pm$ 1.0 MeV of natural parity state $B_s(1^3D_1)$ is within the errorbar of experimental measurement 66 $\pm$ 18 $\pm$ 21 MeV of $B_s(6114)$. We expect more experimental efforts in the future to identify its nature properly.}

However, such a comparison is not necessarily valid because heavier excited meson resonances might decay to alternative final states, such as those involving higher light pseudoscalar multiplicities, light vector mesons, and lower-lying excited heavy mesons \cite{godfrey2016b,sun2014higher,lu2016excited,yu2019analysis,zhong2008strong,campanella2018excited}. The effective heavy meson chiral Lagrangians reported in Ref. \cite{gandhi2019strong} do not contain such extra decay modes.        

\section{Summary}
\label{Summary}

We have carried out a systematic study of the excited $B$ and $B_s$ mesons in the framework of heavy quark effective theory. The masses and the strong two-body decays of excited $B$ and $B_s$ mesons are calculated. Our predictions listed with other results obtained from the various potential model as well as available experimental measurements give the mass range of these mesons. With these predictions, we assign the possible quantum states for the experimentally observed excited $B$ and $B_s$ mesons such as $B_{J}^*(5732)$, $B_{1}(5721)$, $B_{s1}(5830)$, $B_{2}^*(5747)$, $B_{J}(5840)$, $B_{s2}^*(5840)$, $B_{J}(5970)$, and $B_s(6063)$. Furthermore, the strong decay behaviors of the excited $B$ and $B_s$ mesons are investigated with the help of couplings $g_H$, $g_S$, and $g_T$ obtained by P. Colangelo et al. \cite{colangelo2012new} in the strong decay analysis of excited charmed mesons. The $B_{J}^*(5732)$ is identified as $B(1^3P_0)$. The $B_{1}(5721)$ and $B_{2}^*(5747)$ are classified into $P$-wave spin doublet $(1^3P_1, 1^3P_2)$, and the $B_{J}(5840)$ is interpreted as $2^3S_1$. The $B_{s1}(5830)$ and $B_{s2}^*(5840)$ is found to be a strange partners of $B_{1}(5721)$ and $B_{2}^*(5747)$. The branching ratio $\frac{B\pi}{B^*\pi}$ can explain $B_{J}(5970)$ as $1^3D_1$. {\bf We tentatively identify $B_s(6063)$ as $B_s(2^3S_1)$ from the spectroscopy study and $B_s(6114)$ as $B_s(1^3D_1)$ from the strong decay analysis}. In this regard, we expect more information from future experimental studies. The properties of other excited $B$ and $B_s$ mesons are also predicted, which will be useful in future experimental searches.

\section*{Acknowledgment}
\label{Acknowledgment}

{\bf We thank Prof. T. Matsuki for the continuous support throughout this work and for providing critical remarks for the improvement of the paper. Dr. K. Gandhi thank Dr. N. R. Soni for some help in the  uncertainty computation.}

\section*{Appendix: Uncertainty computation}
\label{Appendix: Propagation of Uncertainty}

\textbf{Uncertainty arises in presented results are because of the uncertainties in the various experimental inputs. For the computation of uncertainties in our results, we use the most general technique.  For instance,  the uncertainty in the computation of spin-averaged mass $\bar{M}(a, b)$ because of the uncertainty in the experimental inputs $a$ and $b$ can be written as}
\begin{equation}
\label{eq37} 
\Delta \bar{M}(a, b) = \sqrt{{\left(\frac{\partial \bar{M}(a, b)}{\partial a} {\delta a} \right)}^2 + {\left(\frac{\partial \bar{M}(a, b)}{\partial b} {\delta b} \right)}^2},
\end{equation}
{\bf with $\delta a$ and $\delta b$ are the uncertainties in the $a$ and $b$ respectively.
The uncertainties in the mass splitting between the excited and the low-lying negative parity doublets, hyperfine mass splitting between the members of the doublets, and the strong decay rates are also extracted using the same method.}

\end{document}